\documentclass[5p,numbers,sort&compress,preprint]{elsarticle}
\usepackage[T1]{fontenc}
\usepackage[latin9]{inputenc}
\usepackage{float}
\usepackage{units}
\usepackage{textcomp}
\usepackage{graphicx}

\makeatletter

\providecommand{\tabularnewline}{\\}

\usepackage{hyperref}

\usepackage[english]{babel}

\usepackage[switch, modulo]{lineno}

\pdfoutput=1

\makeatother

\begin{document}

\begin{frontmatter}{}

\title{Progress of the Development of the ELI-NP GBS High Level Applications
}

\author[cor1]{G. Campogiani$^{(a)}$\corref{cor1}, A. Giribono$^{(a)}$, S. Pioli
$^{(a)}$, A. Mostacci$^{(a)}$, L. Palumbo$^{(a)}$}

\author{S. Guiducci $^{(b)}$, G. Di Pirro $^{(b)}$, A. Falone $^{(b)}$,
C. Vaccarezza $^{(b)}$ , A. Variola $^{(b)}$}

\author{S. Di Mitri $^{(c)}$, G. Gaio$^{(c)}$}

\author{J. Corbett$^{(d)}$}

\author{L. Sabato $^{(e)}$}

\author{P. Arpaia$^{(f)}$ }

\author{I. Chaikovska$^{(g)}$}

\cortext[cor1]{giovanna.campogiani@uniroma1.it}

\address{$^{(a)}$Dept. SBAI \textquotedblleft La Sapienza\textquotedblright{}
University, Via Antonio Scarpa,14 00161 Rome, Italy and INFN-Roma1,
Piazzale Aldo Moro,2 00161 Rome, Italy}

\address{$^{(b)}$INFN-LNF, Via Enrico Fermi,40 00044 Frascati, Rome, Italy}

\address{$^{(c)}$Elettra-Sincrotrone Trieste S.C.p.A., Basovizza, Italy}

\address{$^{(d)}$Stanford Synchrotron Radiation Laboratory, Stanford Linear
Accelerator Center Stanford University, Stanford, CA 94309}

\address{$^{(e)}$University of Sannio, Dept. of Engineering, Corso G. Garibaldi,
107, Benevento, Italy}

\address{$^{(f)}$Federico II University of Naples, Dept. of Electrical Engineering
and Information Technology, Via Claudio 21, Naples, Italy}

\address{$^{(g)}$Universit Paris-Sud, LAL,Paris, France}

\address{}
\begin{abstract}
The Gamma Beam System (GBS) is a high brightness LINAC to be installed
in Magurele (Bucharest) at the new ELI-NP (Extreme Light Infrastructure
- Nuclear Physics) laboratory. The accelerated electrons, with energies
ranging from 280 to 720 MeV, will collide with a high power laser
to produce tunable high energy photons (0.2-20MeV ) with high intensity
($10^{13}$ photons/s), high brilliance and spectral purity ($0.1\%$BW),
through the Compton backscattering process. This light source will
be open to users for nuclear photonics and nuclear physics advanced
experiments. Tested high level applications will play an important
role in commissioning and operation. In this paper we report the progress
and status of the development of dedicated high level applications.
We also present the results of the test on the FERMI LINAC of the
electron trajectory control method based on Dispersion Free Steering.
\end{abstract}

\end{frontmatter}{}

\section{Introduction}

Any particle accelerator is an ensemble of many different complex
devices and systems. Each subsystem is properly integrated in a control
system, which enables the operator to access the data in read/write
mode. A set of procedures which act on data accessible by the control
system is called \textquotedblleft application\textquotedblright{}
on the machine. A High Level Application (HLA) is a set of automated
commands or operations that perform a specific measurement/characterization
on the machine. We want to have the possibility to test the applications
on a virtual accelerator before the GBS LINAC commissioning\cite{Giribono:IPAC2017-MOPVA016}.
The control system (CS) of the ELI-GBS is based on EPICS, which is
a distributed software framework based on a client-server architecture\cite{Pioli:2016pav}.
The data from devices is put on the EPICS Channel Access (CA) communication
protocol, from which it is accessible by the client services in the
network.

Matlab Middle Layer (MML)\cite{Corbett:2003yh,mmlmanual,Portmann:894827,Terebilo:2001ba}
is the framework chosen to develop the high level applications. MML
consists of a library of Matlab files that enable the development
of control applications for particle accelerators control. An operator
can access Process Variables (PVs) on the machine control system I/O
Controllers (IOC) or on a machine model. MML software development
is aimed to allow accelerator physicists to develop more easily the
required high level functions for a proper operation of the machine.

The most basic MML functions are used to communicate either with the
online machine or with the virtual one. The two main functions for
data access are $getpv$ and $setpv$. Access to the accelerator hardware
is handled through machine-specific routines to acquire and control
database parameters csuch as BPMs, Magnets, RF systems. Depending
on whether the program is in the simulate or on-line mode, the $getpv$
and $setpv$ commands communicate with the MATLAB Accelerator Toolbox
(simulation) or with actual hardware via EPICS CA (on-line). 

\section{eleMML Architecture}

\begin{figure}[H]
\includegraphics[scale=0.5]{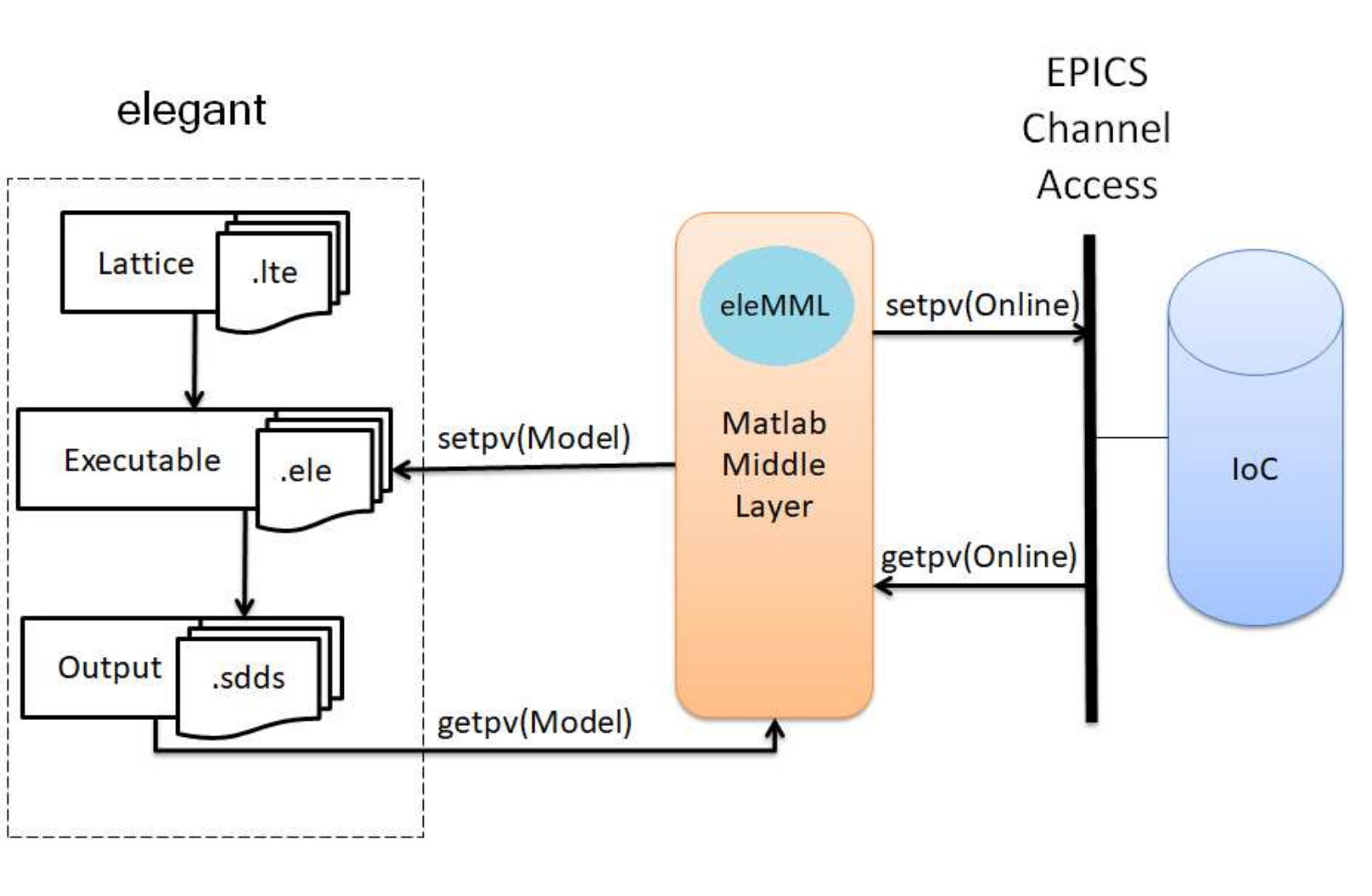}

\caption{Schematic diagram of the architecture of eleMML\label{fig:Schematic-diagram-of-eleMML-1}}
\end{figure}

The integration of the physics engine of $\mathsf{elegant}$\cite{Borland_elegant:a}
with the power of MATLAB enables to develop and test applications
on the ELI-GBS virtual machine, through the Self Describing Data Sets
(SDDS) file protocol\cite{sdds}. The result is eleMML i.e. a set
of software tools developed to interface MML with elegant and to use
the latter as a physics engine for the \textquotedblleft model\textquotedblright{}
operation of High Level Applications. An overview of the architecture
of eleMML is presented in Fig. \ref{fig:Schematic-diagram-of-eleMML-1}. 

The core of eleMML tool-set is the development of a software function
called $\mathsf{sddsReadAny}$ which imports data from sdds-type files
into the Matlab workspace, linking each column or parameter in the
input sdds file into a column array. SDDS Toolkit provides a basic
link between Matlab and elegant through the SDDSMatlab library. This
contains the $\mathsf{sddsload}$ function, which imports an sdds
file into a Matlab structure-type variable. The $\mathsf{sddsReadAny}$
function goes one step further in providing the user directly with
the simulation output data. When a setpv call is made, the simulation
data is updated and element parameters changed. In order to write
data to the different elements in the simulation mode, we exploit
the $\mathsf{-macro}$ option of $\mathsf{elegant}$ call. The $\mathsf{-macro}$
option allows performing text substitutions in the command stream.
Multiple $\mathsf{-macro}$ options may be given. A system call is
made from Matlab when setpv is invoked. 

\begin{figure}[H]
\includegraphics[scale=0.5]{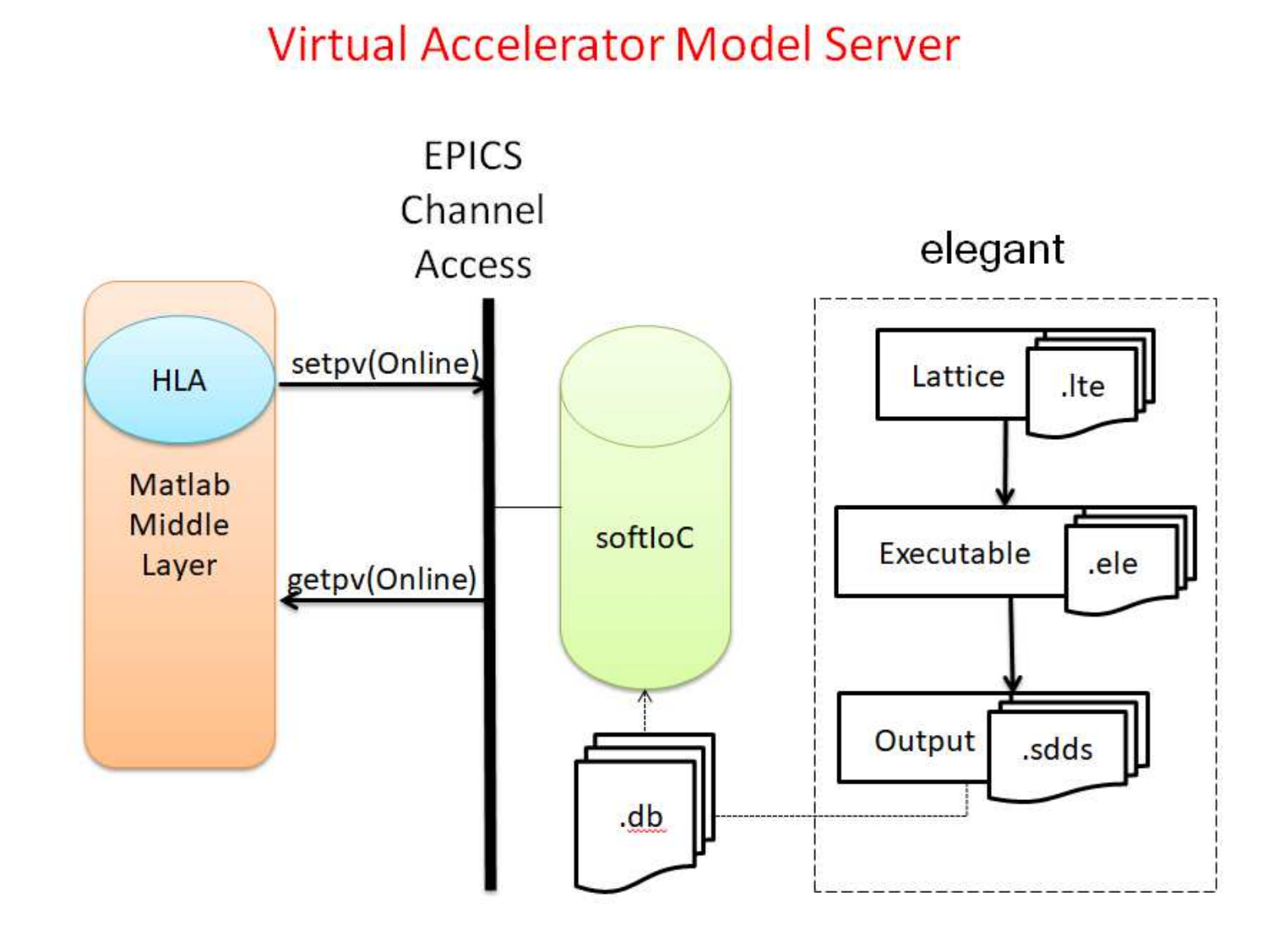}

\caption{\label{fig:va}Virtual test platform that simulates the LINAC response
to HLA commands}
\end{figure}

In preparation for the commissioning and operations of ELI-NP GBS,
a dedicated \textquotedblleft virtual accelerator\textquotedblright{}
(VA) or \textquotedblleft Model Server\textquotedblright{} has been
created. This virtual test platform simulates the LINAC response to
HLA commands. It is based on a soft IOC i.e. a database of process
variable records not associated with real hardware but that can be
used to simulate the behavior of a real device. The soft IOC PVs can
be configured directly from simulated sdds output files, and written
to the CA process variables relative to hardware devices or general
machine, as schematically illustrated in Fig. \ref{fig:va}. 

\section{Simulated bunch length measurement}

One of the first applications needed for commissioning is the bunch
length measurement\cite{Sabato:2016xlu}, for which a model server
was set up to simulate the real measure. Bunch length measurement
procedure is based onthe actual measurement. The bunch length procedure
is based on streaking the beam with a transverse deflecting cavity,
stepping the voltage phase around the $\psi=0\,rad$ value, to measure
the centroid position downstream. A linear fit is performed to correlate
the vertical beam centroid at the screen $C_{y}(\psi)$ with respect
to the phase itself. The slope of the linear fit is the calibration
factor $K_{cal}(\psi)$ that enables to calculate the bunch length.

\begin{figure}[H]
\includegraphics[scale=0.55]{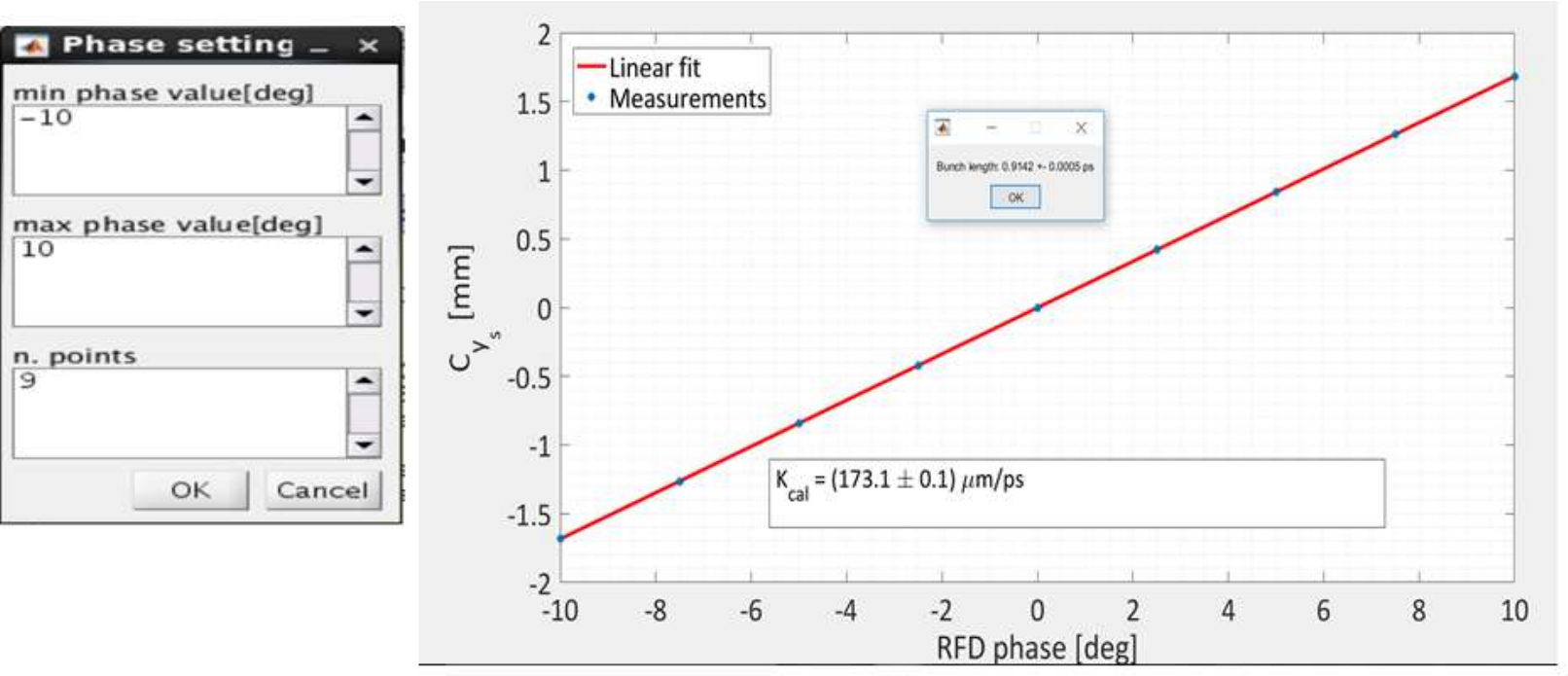}

\caption{\label{fig:Screenshot-of-the}Screenshot of the box in which an operator
can select the parameters to perform the bunch length measurement
(left) and resulting measurement data (right).}
\end{figure}

The results of the elegant simulations have been pushed on the VA
to test the bunch length measurement application. Fig. \ref{fig:Screenshot-of-the}
shows the screenshots of the high level application, with the operator
panel on the left and the measured data on the right. 

\section{Experimental trajectory control at Fermi}

\begin{figure}[H]
\includegraphics[scale=0.37]{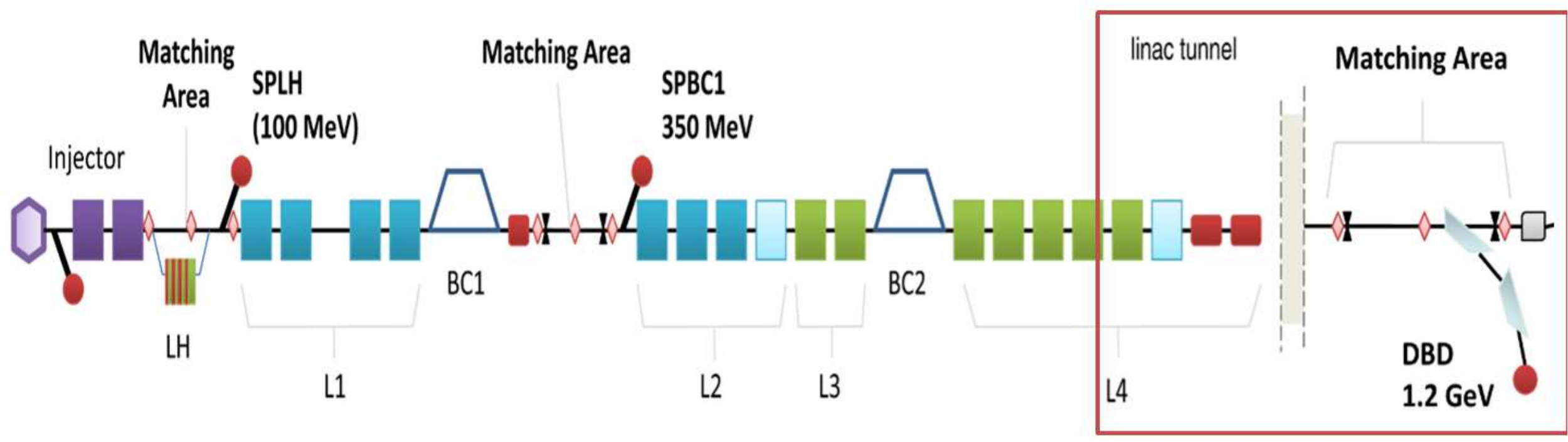}

\caption{a) Schematic layout of the FERMI machine up to the Matching Area preceding
the undulators. The red box highlights the portion of the machine
on which the DFS test was performed. The diagnostic beam dump has
been used to measure the energy of the test beam.\label{fig:Fermi@Elettra-facility-layout-1}}
\end{figure}

Beam trajectory control is important to maintain quality beam at the
IP and subsequent properties of the radiation for the experiments.
As the GBS will operate in a wide range of energies and electron beam
parameter settings, LINAC properties should be optimized for all working
points of the machine. The ideal operation is when the electron beam
trajectory lies on the electromagnetic center of the all the active
elements. There are different methods to control the trajectory and
compensate misalignment, which have been studied for the low energy
line of the ELI-NP GBS. As reported in \cite{Campogiani:2017sug},
Dispersion Free Steering\cite{Pfingstner:2017mhv,Schulte:879699,1991NIMPA.302..191R,Latina:2135837}
enables to optimize the beam trajectory better than other methods.
We made a test of DFS at Fermi, an international FEL facility in Basovizza,
Italy\cite{47c01aa7-f500-431b-8043-aebff5d7b41e,Allaria2013a}. The
acceleration, compression and transport of the electron beams occupies
approximately the first 300 m of the machine. The machine operates
in single bunch mode. The bunch charge was 700 pC and the energy at
the starting point of the correction was 1.2 GeV. We used 5 horizontal
and vertical correctors on a section of the LINAC (red box in Fig.
\ref{fig:Fermi@Elettra-facility-layout-1}) comprising the last accelerating
section and the transport line to the undulators' hall. The relevant
beam parameters are summarized in Table (\ref{tab:Summary-of-machine}).
\begin{center}
\begin{table}[H]
\begin{centering}
\begin{tabular}{|c|c|c|}
\hline 
Electron beam & Value & Units\tabularnewline
\hline 
\hline 
Energy & $\approx$1.3 & GeV\tabularnewline
\hline 
Slice energy spread & < 0.2 & MeV\tabularnewline
\hline 
Charge & 700 & pC\tabularnewline
\hline 
Norm. emittance & 0.8-1 & mm-mrad\tabularnewline
\hline 
\end{tabular}
\par\end{centering}
\caption{\label{tab:Summary-of-machine}Summary of some of the key electron
beam parameters of the working point chosen for the DFS test.}
\end{table}
\par\end{center}

Our HLA tool were ported to the resident TANGO control system through
existing Tango-MATLAB binding. Moreover, specific utilities were developed
online in order to perform the measurement, and coordination with
pre-existing measurement tools was achieved. In particular the Fermi
LINAC has an ongoing Fast Trajectory Feedback always active. We unplugged
the DFS correctors from the active feedback in order to perform our
measurements and tests. The reference trajectory was the initial electron
orbit set by machine operators. We used 5 horizontal and vertical
correctors on a section of the LINAC comprising the last accelerating
section (L04.06) and the transport line to the undulators' hall, corresponding
to the part in the rectangular box of Fig. \ref{fig:Fermi@Elettra-facility-layout-1}.
The portion is called TLS and goes from L4.06 accelerating section
to the end of the transport line to the undulators.

To change the beam energy, the phase of the accelerating field with
respect to the electron beam in the LINAC4 accelerating section L04.06
was varied. The relative phase between the beam and the accelerating
field was varied by 90\textdegree . This leads to a $\Delta E=25MeV$
between the nominal and the offset beam (about 25\% relative energy
difference). This is smaller than the usually introduced energy changes,
which are on the order of 5\% to 10\%. The shot to shot orbit jitter
affects the orbit measurement for the dispersion free steering. An
averaging strategy can improve the measurement, as described in the
following section.

\subsection{Experimental results}

\begin{figure}[H]
\includegraphics[scale=0.37]{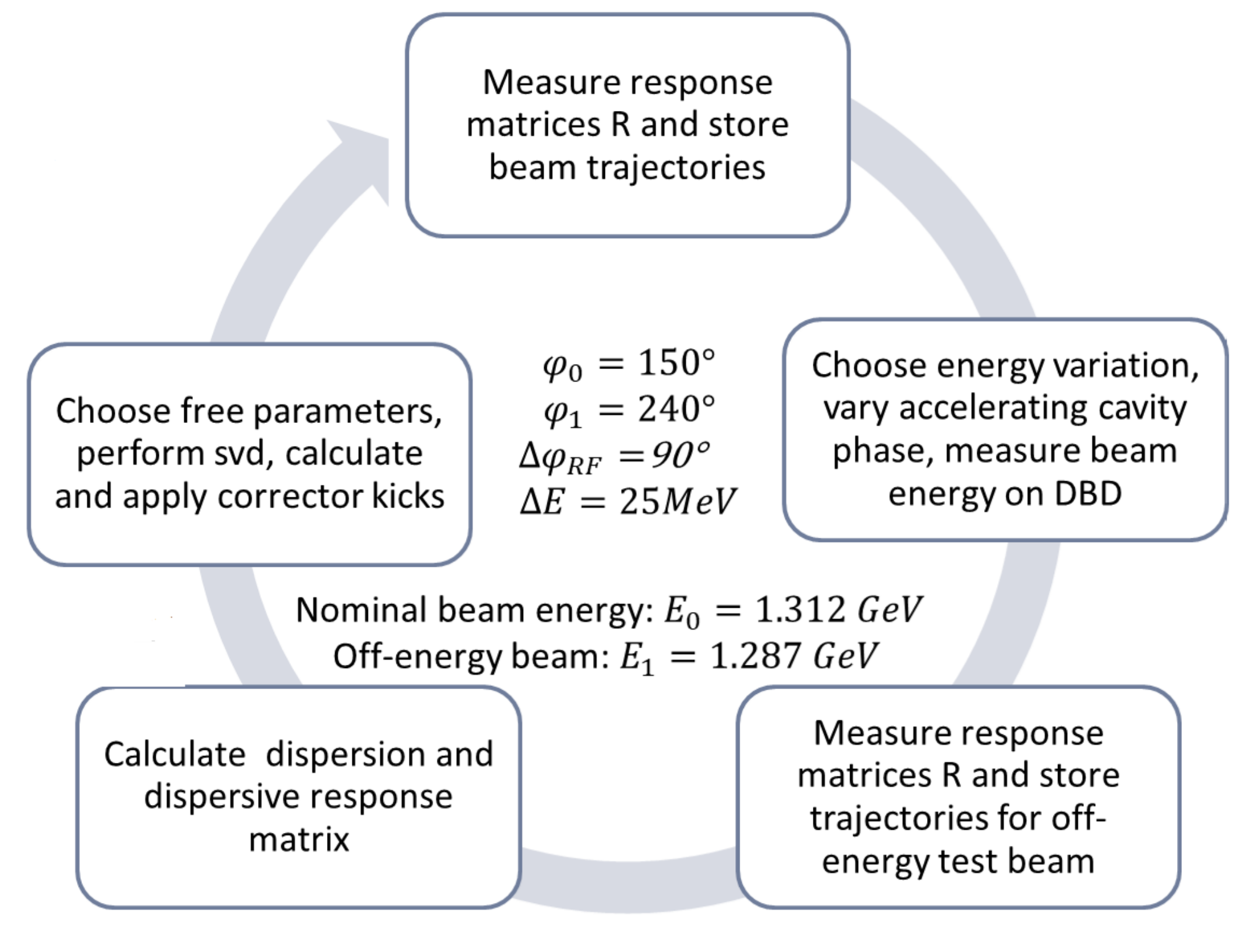}

\caption{\label{fig:dfsconcept-1}Conceptual scheme of the steps needed to
perform DFS correction, with details on the working point of the off-energy
beam chosen to perform our measurement.}
\end{figure}

The experimental procedure steps are summarized in Fig. \ref{fig:dfsconcept-1}.

Dispersion measurement is based on measuring the orbit for different
beam energies which can be expressed as
\begin{equation}
D=\frac{\delta x}{\nicefrac{\Delta E}{E}}=\frac{x(E_{0}(1-\delta))-x(E_{0})}{\nicefrac{\Delta E}{E}}\label{eq:dispersion-eq}
\end{equation}

\begin{figure}[H]
\includegraphics[scale=0.29]{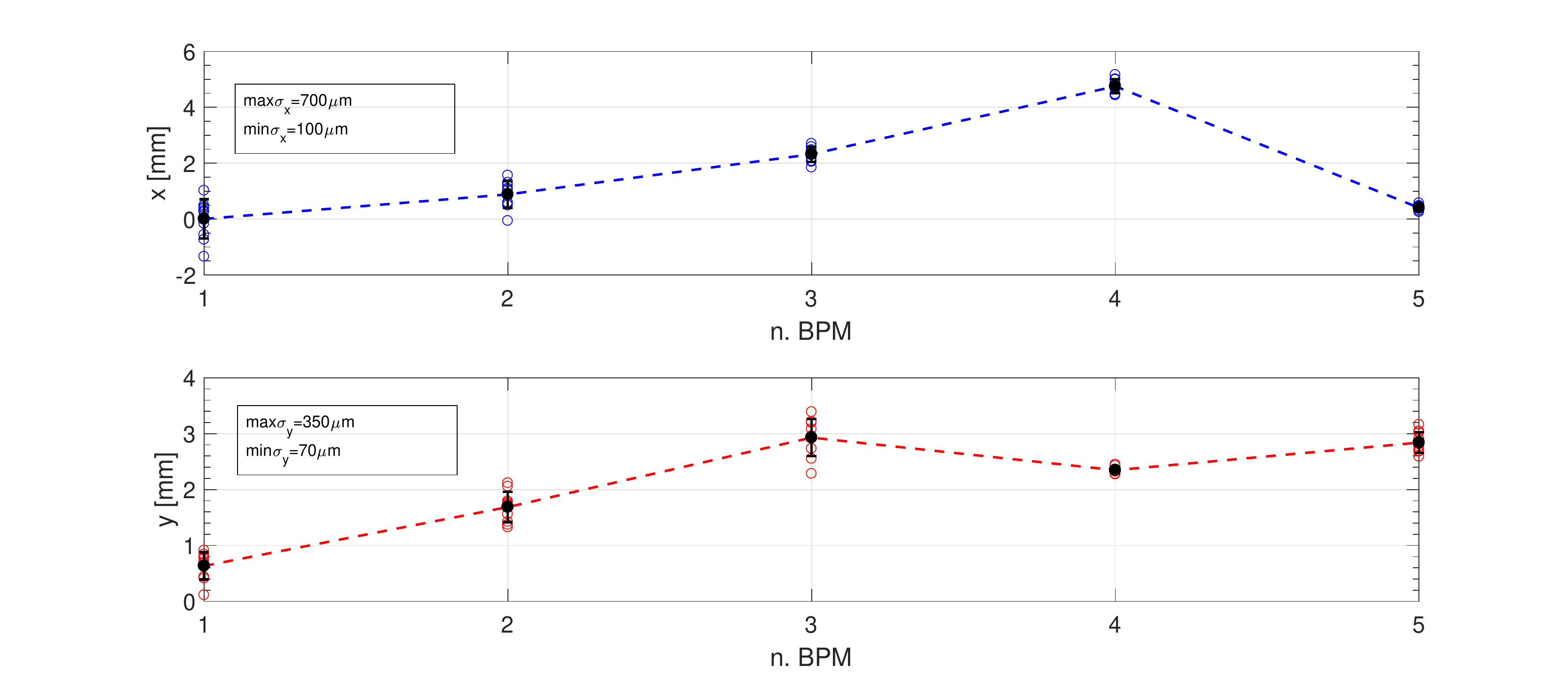}

\caption{Before-correction: horizontal (top) and vertical (bottom) beam trajectory
measurement. Empty dots: values of the position at the BPMs for ten
consecutive trajectory samples acquired. The data shows the fluctuation
due to launching conditions of the beam at the entrance of the LINAC.
Solid line: average value with calculated $\sigma-wide$ error-bars.
The text-box on the plot shows the value of the precision of the measurement
. The single shot resolution of each BPM is $\sigma_{BPM}=5\mu m$.\label{fig:a)-Before-correction:-horizontal-1}}
\end{figure}

In order to achieve a stable measurement, 10 BPM samples were averaged
for each trajectory and dispersion measurement, as shown in Fig. (\ref{fig:a)-Before-correction:-horizontal-1}).

The optimal weighting of the difference orbit with respect to the
nominal orbit was determined to be $\omega=1$0. This is chosen empirically
altough a theoretical estimate can be obtained from
\begin{equation}
\omega=\frac{\sigma_{bpm\,res}^{2}+\sigma_{bpm\,align}^{2}}{2\sigma_{bpm\,res}^{2}}
\end{equation}
where $\sigma_{bpm,res}$ is the BPM precision and $\sigma_{bpm,align}$
is an estimate of the standard deviation of the BPM misalingment distribution.

Fig. (\ref{fig:Horizontal-and-vertical-correctors}) shows the calculated
corrector kicks: the horizontal values are one order of magnitude
higher than the vertical ones, as expected from the response matrix
analysis.
\begin{center}
\begin{figure}[H]
\includegraphics[scale=0.29]{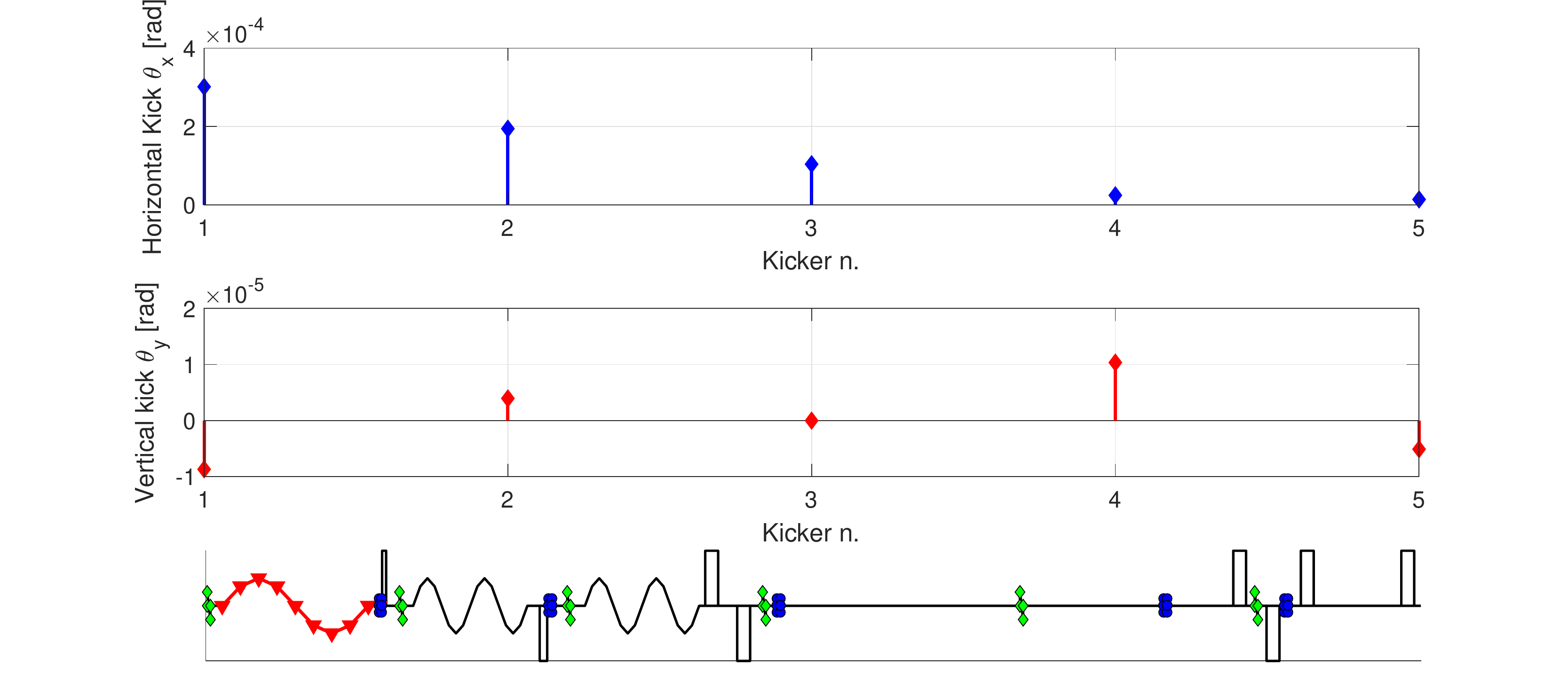}

\caption{\label{fig:Horizontal-and-vertical-correctors}Horizontal(top subplot)
and vertical(bottom subplot) corrector kicks calculated with DFS and
applied to perform the correction.}
\end{figure}
\par\end{center}

\begin{figure}[H]
\raggedright{}\includegraphics{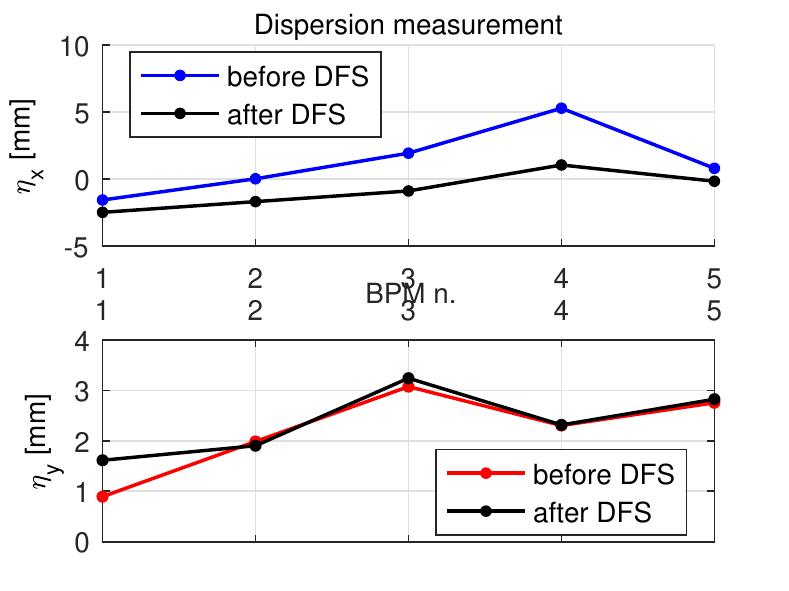}\caption{\label{fig:Measured-dispersion-(averaged}Measured residual dispersion
before and after applying DFS correction, respectively for the horizontal
(top subplot) and vertical (bottom subplot) planes.}
\end{figure}

The result of the DFS correction is summarized in Fig. (\ref{fig:Measured-dispersion-(averaged}),
where the horizontal and vertical dispersion before and after correction
are shown. Table \ref{tab:DFS-params} contains a summary of the electron
beam and working point parameters for the experiment.
\begin{center}
\begin{table}[H]
\begin{centering}
\begin{tabular}{|c|c|c|}
\hline 
Parameter & Value & Unit\tabularnewline
\hline 
\hline 
$E_{0}$ & 1.312 & GeV\tabularnewline
\hline 
$E_{1}$ & 1.287 & GeV\tabularnewline
\hline 
$\nicefrac{\Delta E}{E}$ & 2 & \%\tabularnewline
\hline 
$\omega$ & 10 & \tabularnewline
\hline 
$\beta$ & 1 & \tabularnewline
\hline 
\end{tabular}
\par\end{centering}
\caption{\label{tab:DFS-params}Summary of the free parameter values chosen
for the DFS off-energy correction. $\omega$ is proportional to the
ratio between residual dispersion correction and bpm-zeroing, while
$\beta\protect\neq0$ limits the maximum corrector kick amplitudes. }
\end{table}
\par\end{center}

The horizontal dispersion was corrected significantly while the vertical
correction is almost negligible. An explanation for this comes from
analyzing the values of the residual dispersion before the correction:
in the vertical plane the initial value is lower than for the horizontal
plane, thus the energy difference used for correcting the horizontal
plane was insufficient for the vertical case.

DFS proved to be capable of correcting the trajectory but care had
to be put into defining the free parameters $\omega,\beta$ and the
energy difference $\Delta E$ between nominal and off-energy beam.

\section{Summary}

We have presented the status of the development of high level applications
for ELI-NP GBS. Integrating elegant with MML has lead to the development
of the eleMML architecture. This allows us to test the tools for commissioning
on a virtual accelerator based on EPICS soft-IOCs, such as a transverse
deflecting bunch length measurement application. For a complex application
such as Dispersion Free Steering, we tested the method at Fermi LINAC.
Fermi is a machine routinely used by international users so we didn't
expect to find very high values of residual dispersion as the machine
has been already and is continuously optimized. Nevertheless the DFS
method was successfully validated on an existing machine. With eleMML
framework, collaborative development of more high level applications
for each working group will be possible, enabling the collaboration
to simulate commissioning before the machine comes online. For istance
an application is under development to perform the beam-based alignment
of individual accelerating sections.

\section*{\textemdash \textemdash \textemdash \textemdash \textemdash \textendash{}}

\bibliographystyle{elsarticle-num}
\bibliography{eaac17bibliography}

\end{document}